\documentclass[floatfix,10pt,onecolumn,amsmath,amssymb]{revtex4}

\usepackage{epsf}
\usepackage{graphicx}  % Include figure files
\usepackage{dcolumn}   % Align table columns on decimal point
\usepackage{bm}        % bold math
\usepackage{color}

\newcommand{\be}{\begin{equation}}
\newcommand{\en}{\end{equation}}
 \newcommand{\bea}{\begin{eqnarray}}
 \newcommand{\ena}{\end{eqnarray}}

\begin{document}

\title{\boldmath Universality in the stress tensor for Holographic fluids at the finite cutoff surface via fluid/gravity correspondence}

 \author{Ya-Peng Hu$^{1,2,5~}$\footnote{Electronic address: huyp@nuaa.edu.cn}  Xin-Meng Wu $^{1,~}$\footnote{Electronic address: wuxm@nuaa.edu.cn}  Jun-Peng Hou $^{4~}$ \footnote{Electronic address: junpeng.hou@utdallas.edu} Hongsheng Zhang $^{3,5~}$ \footnote{Electronic address: sps\_zhanghs@ujn.edu.cn}, }
\affiliation{$^1$ College of Science, Nanjing University of Aeronautics and Astronautics, Nanjing 210016, China\\
 $^2$ Instituut-Lorentz for Theoretical Physics, Leiden University, Niels Bohrweg 2, Leiden 2333 CA, The Netherlands\\
 $^3$ School of Physics and Technology, University of Jinan, 336 West Road of Nan Xinzhuang, Jinan, Shandong 250022, China\\
 $^4$ Department of Physics, The University of Texas at Dallas, Richardson, Texas 75080-3021, USA\\
 $^5$ State Key Laboratory of Theoretical Physics, Institute of Theoretical Physics, Chinese Academy of Sciences, Beijing, 100190, China}

%\date{ \today}
%% %simple case: 2 authors, same institution
%% \author{A. Uthor}
%% \author{and A. Nother Author}
%% \affiliation{Institution,\\Address, Country}

% more complex case: 4 authors, 3 institutions, 2 footnotes
%\author[a,b,c,1]{Ya-Peng Hu,\note{Corresponding author.}}
%\author[c]{S. Econd,}
%\author[a,2]{T. Hird,}%\note{Also at Some University.}}
%\author[a,2]{and Fourth}

% The "\note" macro will give a warning: "Ignoring empty anchor..."
% you can safely ignore it.

%\affiliation[a]{One University,\\some-street, Country}
%\affiliation[b]{Another University,\\different-address, Country}
%\affiliation[c]{A School for Advanced Studies,\\some-location, Country}

%\affiliation[a]{College of Science, Nanjing University of Aeronautics and Astronautics, \\Nanjing 210016, China}
%\affiliation[b]{Instituut-Lorentz for Theoretical Physics, Leiden University, \\Niels Bohrweg 2, Leiden 2333 CA, The Netherlands}
%\affiliation[c]{School of Physics and Technology, University of Jinan, \\336 West Road of Nan Xinzhuang, Jinan, Shandong 250022, China}
%\affiliation[d]{Key Laboratory of Theoretical Physics, Institute of Theoretical Physics, \\Chinese Academy of Sciences, Beijing 100190, China}
%\affiliation[e]{Department of Physics, The University of Texas at Dallas, \\Richardson, Texas 75080-3021, USA}

% e-mail addresses: one for each author, in the same order as the authors
%\emailAdd{huyp@nuaa.edu.cn}
%\emailAdd{wuxm@nuaa.edu.cn}
%\emailAdd{junpeng.hou@utdallas.edu}
%\emailAdd{hongsheng@shnu.edu.cn}

\begin{abstract}
{We investigate the stress tensor for holographic fluids
at the finite cutoff surface through perturbing the Schwarzchild-AdS black brane background to the first order perturbations
in the scenario of fluid/gravity correspondence. We investigate the most general perturbations of the metric
without any gauge fixing. We consider various boundary conditions and demonstrate the properties of the corresponding holographic fluids.
 The critical fact is that the spatial components of the first order stress
tensors of the holographic fluids can be rewritten in a concordant form, which implicates that there is
an underlying universality in the first order stress tensor. We find this universality in the first order stress tensor
for holographic fluids at the finite cutoff surface by an exhaustive investigation of perturbations of the full bulk metric.}

PACS numbers: 04.70.Dy,~11.25.Tq,~04.65.+e

\end{abstract}

\maketitle

\section{Introduction}
\label{sec:intro}

 Over recent decades, the strongly coupled quantum field theory (QFT) attracted many attentions to investigate since it has been found in many physical systems such as the high-Tc superconductor, non-fermi liquid, strange metal and Quark Gluon Plasma(QGP)~\cite{Zaanen:2015oix,Ammon:2015wua,CasalderreySolana:2011us}, etc. Particularly, for the QGP formed in the ultrarelativistic heavy-ion collisions at the Relativistic Heavy Ion Collider (RHIC) and the Large Hadron Collider (LHC), one finds that it thermalizes rapidly after the collision of the ions and comes into local thermal equilibrium state, and then hadronizes when the local temperature goes down to the deconfinement temperature. Furthermore, an interesting and significant observation is that the QGP behaves like an ideal fluid in the local thermal equilibrium regime, i.e. small shear and bulk viscosities~\cite{CasalderreySolana:2011us}. Theoretically, it deserves to study the fluid-alike aspects of  QGP. Unfortunately, the usual method fails at this energy scale. The coupling coefficient of QGP remains strong near the deconfinement temperature in this regime, thus the widely used perturbation quantum chromodynamics (QCD) breaks down, while the Lattice QCD is not well-suited in dealing with real time physics and Lorentzian correlation functions~\cite{Ammon:2015wua,CasalderreySolana:2011us}.

One interesting progress to study the strongly coupled QFTs is the AdS/CFT correspondence~\cite{Maldacena:1997re,Gubser:1998bc,Witten:1998qj,Aharony:1999ti}. In this scenario, one can obtain the properties of the strongly coupled field theory through investigating a gravitational theory in a higher dimensional bulk spacetime, which is asymptotical to anti-de Sitter spacetimes. The early work showed that the AdS/CFT correspondence can describe the hydrodynamical behavior of the QGP via the dual gravity in the bulk, and also presents the shear viscosity to entropy density ratio $\eta/s$~\cite{Policastro:2001yc}. The ratio takes a universal value $1/4\pi$ for all QFTs in frame of the Einstein gravity~\cite{Buchel:2003tz}. This fact leads to a conjecture that there is a lower bound $\eta/s>1/4\pi$~~\cite{Kovtun:2003wp,Kovtun:2004de}, which is supported both by energy-time uncertainty principle arguments and QGP experimental data~\cite{Song:2010mg}. AdS/CFT correspondence has provided several new insights to investigate the strongly coupled field theory from the dual classical gravitational theory~\cite{Herzog:2009xv,CasalderreySolana:2011us,Policastro:2001yc,Kovtun:2003wp,Buchel:2003tz,Kovtun:2004de}. Since the hydrodynamics can be viewed as an effective description of an interacting quantum field theory in the long wavelength limit, the study of the hydrodynamics via dual gravity has been further developed as the fluid/gravity correspondence \cite{Bhattacharyya:2008jc,Rangamani:2009xk}. This correspondence provides a clear scenario about the correspondence between the boundary fluid and the bulk gravity. One can obtain some detailed information of the holographic fluids, i.e. the stress tensor or charged current of the holographic fluid from the bulk gravity solution~\cite{Bhattacharyya:2008jc,Rangamani:2009xk,Hur:2008tq, Erdmenger:2008rm, Banerjee:2008th,
Son:2009tf, Tan:2009yg, Torabian:2009qk, Hu:2010sn, Hu:2011ze,
Kalaydzhyan:2011vx, Amado:2011zx}.

In the AdS/CFT correspondence or fluid/gravity correspondence, the dual field theory
usually resides on the infinite boundary (conformal boundary or UV boundary) with conformal symmetry~\cite{Bhattacharyya:2008jc,Rangamani:2009xk,Hur:2008tq, Erdmenger:2008rm, Banerjee:2008th,
Son:2009tf, Tan:2009yg, Torabian:2009qk, Hu:2010sn, Hu:2011ze,
Kalaydzhyan:2011vx, Amado:2011zx}. The real fluids do not necessarily have conformal symmetry. Moreover, the QGP found in the RHIC has non-zero shear and bulk viscosities. Hence the AdS/CFT correspondence should be generalized to study the non-conformal fluids. A realization is to break the conformal symmetry by introducing a finite cutoff at the radial direction in the bulk~\cite{Bredberg:2010ky}. It has been found that a Navier-Stokes (NS) fluid can live on the cutoff surface $r=r_c$~\cite{Bredberg:2010ky,Bredberg:2011jq,Compere:2011dx,Cai:2011xv,Niu:2011gu,Compere:2012mt,Eling:2012ni}. In addition, since the radial direction of the bulk spacetime corresponds to the energy scale of the dual field theory~\cite{Balasubramanian:1998de,Susskind:1998dq,Akhmedov:1998vf,de
Boer:1999xf,Kuperstein:2011fn,Heemskerk:2010hk,Faulkner:2010jy,Iqbal:2008by,Sin:2011yh}, the infinite boundary corresponds to
the UV fixed point of the dual field theory, and hence cannot be reached by experiments. From the renormalization group (RG) viewpoint, the dependence of transport coefficients of holographic fluid on the cutoff surface $r_c$ can be interpreted as RG flow. There are several
investigations on the holographic fluids at a finite cutoff surface \cite{Bredberg:2010ky,Bai:2012ci,Hu:2013lua,Hu:2014wka,Cai:2011xv,Kuperstein:2011fn,Brattan:2011my,Camps:2010br,Emparan:2012be,Emparan:2013ila}. It is found that
the holographic fluids on the cutoff surface are usually non-conformal, which is expected and consistent with the fact
that the conformal symmetry has been broken with a finite radial coordinate in the bulk~\cite{Brattan:2011my,Camps:2010br,Emparan:2012be,Emparan:2013ila,Bai:2012ci,Hu:2013lua,Hu:2014wka,Grozdanov:2011aa}. In view of these approaches, the dual fluids at finite cut-off surfaces correspond to more realistic fluids in nature.

In this paper, we will make an exhaustive investigation for the stress tensor at the first order holographic fluids at finite cutoff surfaces in the scenario of fluid/gravity correspondence. We  consider the Schwarzchild-AdS black brane as the background. Note that, in derivation of the first order stress tensor of holographic fluid, one usually chooses some gauge and boundary conditions for the correction terms in the metric perturbations, for examples,  background gauge and the Dirichlet boundary conditions. We also first consider different gauge and boundary conditions in two special cases to obtain the first order stress tensor of holographic fluid at the finite cutoff surface to find some clues for our universal relation of the stress tensor. It should be pointed out that usually different boundary conditions corresponds to the underlying different holographic fluids. Our results show that stress tensors of holographic fluids are the same, which implicates that there is an underlying universality of the stress tensor. In order to clearly demonstrate this universality, we consider the full correction terms in the metric perturbations, and directly investigate the general first order stress tensor without any gauge fixing or  special boundary conditions. After making the comparison with the tensor $W_{AB}$ related to the perturbative equations, we find that there is an underlying relationship between the stress tensor and $W_{AB}$. From this relationship, we find out the underlying universality in the stress tensor, which is independent of any special gauge or boundary conditions.

The paper is organized as follows. In Sec.~II, we focus on the first order perturbative solution for the Schwarzchild-AdS black brane in the scenario of fluid/gravity correspondence. In Sec.~III, we obtain the general first order stress tensor of holographic fluid at the finite cutoff surface, where the full correction terms in the metric perturbations are considered. In Sec.~IV, the explicit stress tensor is obtain in two special cases, where different gauge and boundary conditions are adopted. An interesting result is that the spatial components of the first order stress tensor can be written in the same form, which implicates that there is some universality in the stress tensor. In Sec.~IV, the underlying universality in the first order stress tensor of holographic fluid is derived in the perturbative Schwarzchild-AdS black brane, and the result is used to compared with the tensor $W_{AB}$. Sec.~VI is devoted to the conclusion and discussion.

\section{The first order perturbative equations of the Schwarzschild-AdS black brane solution}
The action of the five dimensional spacetime with a negative cosmological constant $\Lambda=-6/\ell^2$ in the Einstein gravity is
\begin{equation}
\label{IVaction1} I=\frac{1}{16 \pi G}\int_\mathcal{M}~d^5x
\sqrt{-g^{(5)}} \left(R-2 \Lambda
\right),
\end{equation}
while the corresponding equation of motion is
\begin{eqnarray}
\label{IVeqs1}
R_{AB } -\frac{1}{2}Rg_{AB}+\Lambda g_{AB}&=&0~,
\end{eqnarray}
where $\ell$ represents the AdS radius, and $\ell=1$ and $16\pi G=1$ have been assumed for the later calculation convenience.

The Schwarzschild-AdS black brane solution is obtained from the above equation as
\begin{eqnarray}
ds^2=\frac{dr^2}{r^2f(r)}+r^2
 \left(\mathop\sum_{i=1}^{3}dx_i^2 \right)-r^2f(r) dt^2, \label{IVSolution}
\end{eqnarray}%
and
\begin{eqnarray}
\label{IVf-BH} f(r) &=& 1-\frac{2M}{r^{4}}.
\end{eqnarray}%
In order to avoid the coordinate singularity, we use the Eddington-Finkelstin coordinates, and hence the Schwarzschild-AdS black brane solution reads,
\begin{eqnarray}\label{IVSolution1}
ds^2 &=& - r^2 f(r)dv^2 + 2 dv dr + r^2(dx^2 +dy^2 +dz^2),
\end{eqnarray}
where $v=t+r_*$, and $r_*$ is the tortoise coordinate,
$dr_*=dr/(r^2f)$.

Note that, we will investigate the holographic fluids at a finite cutoff hypersurface $r=r_c$ ($r_c$ is a constant), and the fluid usually resides on the flat spacetime like $ds^2=-dv^2+dx^2+dy^2 +dz^2$. Therefore, we should make a further coordinate transformation $v\rightarrow v/\sqrt{r_c^2 f(r_c)}$ and $x_i\rightarrow x_i/r_c$, and hence the Schwarschild-AdS black brane solution becomes
\begin{eqnarray}\label{IVSolution2}
ds^2 &=& - \frac{r^2 f(r)}{r_c^2 f(r_c)}dv^2 + \frac{2}{r_c \sqrt{f(r_c)}} dv dr + \frac{r^2}{r_c^2}(dx^2 +dy^2 +dz^2).
\end{eqnarray}
Then, we boost the above static black brane with a constant velocity at the cutoff surface, which leads to the five-dimensional boosted Schwarzschild-AdS black brane solution
\begin{eqnarray}   \label{IVrnboost}
ds^2 &=& - \frac{r^2 f(r)}{r_c^2 f(r_c)}( u_\mu dx^\mu )^2 - \frac{2}{r_c \sqrt{f(r_c)}} u_\mu dx^\mu dr + \frac{r^2}{r_c^2} P_{\mu \nu} dx^\mu dx^\nu,  ~~
\end{eqnarray}
with
\begin{equation}
u^v = \frac{1}{ \sqrt{1 - \beta_i^2} },~~u^i = \frac{\beta_i}{
\sqrt{1 - \beta_i^2} },~~P_{\mu \nu}= \eta_{\mu\nu} + u_\mu u_\nu\ .
\label{IVvelocity}
\end{equation}
where $x^\mu=(v,x_{i})$ are the boundary coordinates at the finite cutoff surface, $P^{\mu\nu}=\eta^{\mu\nu}+u^\mu u^\nu$ is the projector onto spatial directions, and the constants $\beta^i$ are velocities. The metric~(\ref{IVrnboost}) describes a uniformly boosted black brane moving at the velocity $\beta^i$, and the dual holographic fluid at the finite cutoff surface is just the ideal fluid with the same constant velocity.

 To obtain the transport coefficients of the holographic fluids, we need to perturb the solution (\ref{IVrnboost}). In the scenario of fluid/gravity correspondence, a natural and simple way is to lift the constant parameters into the functions of the boundary coordinates $x^\mu$, i.e., $\beta^i(x^\mu)$ and $M(x^\mu)$. Therefore, ~(\ref{IVrnboost}) will be no longer the solution of the field equation (\ref{IVeqs1}), and hence extra correction terms are required to be introduced to make (\ref{IVrnboost}) be a self-consistent solution. Moreover, for the first order perturbative solution, one can first obtain the extra correction terms at the origin $x^{\mu}=0$. Then, considering the SO(3) symmetry on the boundary, one can obtain the extra correction terms at any point $x^{\mu}$, i.e. making the extra correction terms at the origin $x^{\mu}=0$ into a covariant form \cite{Bhattacharyya:2008jc,Rangamani:2009xk,Hur:2008tq}. In detail, the parameters around $x^\mu=0$ expanded to the first order are
\begin{eqnarray}
\beta_i(x^\mu)&=&\partial_{\mu} \beta_{i}|_{x^\mu=0}
x^{\mu},~M(x^\mu)=M(0)+\partial_{\mu}
 M|_{x^\mu=0} x^{\mu}, \label{IVExpand}
\end{eqnarray}
where $\beta_i(0)=0$ have been assumed at the origin $x^{\mu}=0$, and the first order extra correction terms around $x^\mu=0$ are
\begin{center}
\begin{equation}
\begin{split}
ds_{(1)}^2 =& \frac{ k(r)}{r^2}dv^2 +
2\frac{h(r)}{r_c \sqrt{f(r_c)}}dv dr + 2 \frac{j_i(r)}{r^2}dv dx^i+\frac{B(r)}{r^2}dr^2 +2\frac{B_i(r)}{r^2}drdx^i
\\&+\frac{r^2}{r_c^2} \left(\alpha_{ij}(r) -\frac{2}{3} h(r)\delta_{ij}\right)dx^i dx^j,
 \label{IVcorrection}
\end{split}
\end{equation}
\end{center}
Usually the extra correction terms (\ref{IVcorrection}) can be simplified by choosing some gauge, i.e. the background field gauge in \cite{Bhattacharyya:2008jc}. Here we just use the full correction terms of the metric perturbations without any gauge fixing. Then we choose more boundary conditions after taking different gauge conditions into account, and hence we are more convenient to see the universality of the stress tensor for the first order holographic fluids. A useful tensor is defined as follows
\begin{eqnarray}
&&W_{AB} = R_{AB} + 4g_{AB}, \label{IVTensors1}
\end{eqnarray}
which is in fact related to the gravitational equation. After inserting the metric (\ref{IVrnboost}) with non-constant parameters and (\ref{IVExpand}) into $W_{AB }$ , the nonzero $-W_{AB }$ is usually considered as the first order source terms $S^{(1)}_{AB }$, while the first order perturbation solution around $x^\mu=0$ can be obtained from the vanishing $W_{AB} = (\text{effect from correction})- S^{(1)}_{AB }$, which are casted in the appendix~\ref{A}. Here, the ``effect from correction" means the correction to $W_{AB}$ from (\ref{IVcorrection}).

\section{The first order stress tensor of holographic fluid at the finite cutoff surface}

According to the AdS/CFT correspondence, the stress tensor of the holographic fluids residing at the finite cutoff surface is~\cite{Bredberg:2010ky}
\begin{equation}
T_{\mu\nu}=2(K_{\mu\nu}-K\gamma_{\mu\nu}-C\gamma_{\mu\nu}) \label{Stress tensor}
\end{equation}
where $\gamma_{\mu\nu}$ represent the boundary metric obtained from the  ADM decomposition
\begin{eqnarray}
ds^2 = \gamma_{\mu\nu}(dx^\mu + V^\mu dr)(dx^\nu + V^\nu dr) + N^2
dr^2\ ,
\end{eqnarray}
the extrinsic curvature $K$ is given by $K_{\mu\nu}=-\frac{1}{2}(\nabla_{\mu}n_{\nu}+\nabla_{\nu}n_{\mu})$, and $n^{\mu}$ is the unit normal vector pointing toward the $r$ increasing direction on a constant hypersurface $r=r_c$. The last term in (\ref{Stress tensor}) is usually added to cancel the divergence of the stress tensor in the case with infinite boundary, i.e. $r_c$ to the infinity~\cite{Balasubramanian:1999re,de Haro:2000xn,Emparan:1999pm}. In the finite cutoff surface case, there is in fact no divergence of the stress tensor. The simple reason that we still add it in the stress tensor is that it should has a proper limit at the infinite boundary with $C=3$~\cite{Bhattacharyya:2008jc,Rangamani:2009xk,Hur:2008tq,Bai:2012ci,Hu:2013lua}, and hence we also often set $C=3$ in (\ref{Stress tensor}) and in the following.

  By using the extra correction terms (\ref{IVcorrection}) and  the metric (\ref{IVrnboost}) with the first order parameters expanded, we obtain the stress tensor of the holographic fluids of zeroth order at the finite cutoff surface
\bea
    T^{(0)}_{v v} &=& 2\left(C-3 \sqrt{f\left(r_c\right)}\right),\nonumber\\
    T^{(0)}_{i i} &=&\frac{-4 M+2\left(3-C \sqrt{f\left(r_c\right)}\right) r_c^4}{\sqrt{f\left(r_c\right)} r_c^4}, \label{STBackground}
\ena
and the first order stress tensor
\bea
    T^{(1)}_{v v} &=&-2 \partial_i \beta_i + \frac{\left(-2 C+ 9 \sqrt{f\left(r_c\right)}\right) k\left(r_c\right)}{ r_c^2} + 6 \sqrt{f(r_c)} h(r_c) + 3 f(r_c) \sqrt{f(r_c)} B(r_c) \nonumber\\& &+ 2 r_c \sqrt{f(r_c)} h'(r_c) - r_c \sqrt{f(r_c)} F(r_c),\nonumber\\
    T^{(1)}_{v i} &=& \frac{\partial_i M}{f(r_c) r_c^4} - \partial_v \beta_i + \frac{2 \left(2-C \sqrt{f\left(r_c\right)}+3 f\left(r_c\right)\right)  j_i\left(r_c\right)}{\sqrt{f\left(r_c\right)} r_c^2}-\frac{\sqrt{f\left(r_c\right)} j_i'\left(r_c\right)}{ r_c},\nonumber\\
    T^{(1)}_{i j} &=&-2\partial_{(i}\beta_{j)}+2\delta _{i j}\frac{\partial_v M}{f\left(r_c\right) r_c^4}-2\left(C+\frac{2 M-3 r_c^4}{\sqrt{f\left(r_c\right)} r_c^4}\right)\alpha_{i j}\left(r_c\right)-\sqrt{f\left(r_c\right)} r_c \alpha_{i j}'\left(r_c\right)\nonumber\\& &\nonumber\\
    & & +2\delta _{i j}\left( \left(\frac{2}{3}C + \frac{5}{3}\frac{2M - 3 r_c^4}{rc^4 \sqrt{f(r_c)}} \right) h(r_c) - \frac{2 r_c \sqrt{f(r_c)}}{3} h'(r_c) + \frac{\sqrt{f(r_c)}(2M-3r_c^4)}{2 r_c^4} B(r_c)\right)\nonumber\\
    & & +2\delta _{i j}\left( \frac{2M+r_c^4}{2 r_c^6 \sqrt{f(r_c)}} k(r_c) - \frac{\sqrt{f(r_c)}}{2 r_c} k'(r_c) \right) + \delta_{ij} \sqrt{f\left(r_c\right)} r_c F(r_c) + 2 \delta_{ij} \partial_k \beta_k.\label{BoundaryST}
\ena
where a function $F(r)$ appears to relate to the trace of the tensor perturbation modes, i.e., $F(r)\equiv \sum_i\alpha _{i i}'(r)$. $T^{(1)}_{v i}$ are the momentums of the holographic fluids, which are related to the vector perturbation $j_i(r)$. Usually, they are assumed to be zero by choosing suitable boundary conditions $j_i(r)$, which corresponds to a co-moving frame. We will work in this frame in the followings.

%%%%%%%%%%%%%%%%%%%%%%%%%%%%%%%%%%%%%%%%%%%%%%%%%%%%%%%%%%%%%%%%%%%%%%%%%%%%%%%%%%%%%%%%%%%%%%%%%
%%%%%%%%%%%%%%%%%%%%%%%%%%%%%%%%%%%%%%%%%%%%%%%%%%%%%%%%%%%%%%%%%%%%%%%%%%%%%%%%%%%%%%%%%%%%%%%%%

\section{two heuristic examples for the universality }

Before demonstrating the universal relations in the stress tensor of the holographic fluids, it may be beneficial to study some  naive examples.
In the holographic fluid studies,  we can choose some gauge and boundary conditions to solve the Einstein equations, and hence find out the explicit stress tensors to read off the corresponding quantities, for example, the energy density $\rho$, pressure $p$ and viscosities. Now one can explicitly solve the first order correction terms by choosing the so-called background gauge conditions ($G$ represents the full metric)
\begin{equation}
G_{rr}=0,~~G_{r\mu}\propto u_{\mu},~~Tr((G^{(0)})^{-1}G^{(1)})=0.\label{gauge}
\end{equation}
with the Dirichlet boundary conditions in Landau frame ~\cite{Bai:2012ci}. Different boundary conditions correspond to different underlying physics, which usually deduces different stress tensors of the fluids. Therefore, we will investigate several special cases in this section by choosing different boundary conditions. Generally different gauge conditions should not affect the stress tensor in principle. However, since the boundary conditions under different gauge conditions can be different, thus the stress tensor usually becomes also different. For the five dimensional spacetime, there are five gauge freedoms for the correction terms in (\ref{IVcorrection}). Since the correction terms $B_i(r)$ do not appear in the first order perturbative equations, $B_i(r)=0$ or $G_{rr}=0$ can be assumed, which means that three gauge freedoms are fixed. There are two residue gauge freedoms. For simplicity, we consider two special cases with different gauge conditions in the followings. For the same gauge conditions, we also investigate different boundary conditions.

\subsection{The First Case : $F(r)=0$ and $h(r)=0$ }
Note that, in our paper, we concentrate on the most general correction terms in (\ref{IVcorrection}). That is, not only the full metric perturbations have been considered, but also the tensor perturbation modes can be non-traceless, i.e. $\sum_i\alpha _{i i}(r)\neq 0$ . Moreover, there is an interesting result obtained from the first order perturbative equations in the appendix~\ref{A}, where a function $F(r)$ appears to relate to the trace of the tensor perturbation modes, i.e., $F(r)\equiv \sum_i\alpha _{i i}'(r)$. In the first naive example we impose two boundary conditions $F(r)=0$ and $h(r)=0$, which are apparently different from the background gauge conditions in (\ref{gauge}).

By using these gauge conditions, the $\alpha_{ij}(r)$ can be solved
\begin{equation}
\alpha_{ij}(r)=\alpha(r)\left\{(\partial_i \beta_j + \partial_j
\beta_i )-\frac{2}{3} \delta_{ij}\partial_k \beta^k \right\} +b \delta_{ij},
\end{equation}
where $b$ is a constant and
\begin{equation}
\alpha(r)= r_c\sqrt{f(r_c)} \int_{r_c }^{r}\frac{s^{3}-r_{+}^3}{-s^{5}f(s)}ds.
\end{equation}
Substituting it into $W_{rr}$ and $W_{ii}$, we obtain the solutions of $B(r)$ and $k(r)$
\begin{eqnarray}
B(r)&=& \frac{C_B r^4}{r^4 - 2M},\\
k(r)&=& C_k +{\frac{r^2}{2}C_{k1}}- \frac{C_B r^4}{r_c^2 f(r_c)} + \frac{2 \partial_k \beta^k r^3}{3 r_c \sqrt{f(r_c)}}.\label{Krsolution}
\end{eqnarray}
Note that, the parameter $C_{k1}=0$ is the result when one submits (\ref{Krsolution}) into $W_{ii}=0$ in (\ref{A6}). Therefore, there are three undetermined parameters $b$, $C_B$ and $C_k$, which needs three boundary conditions to fix them. One can set several different boundary conditions, which correspond to different physical conditions. In the followings, we impose two boundary conditions. The two sets of boundary conditions are $\alpha_{ij}(r_c)=0$, $k(r_c)=0$, $B(r_c)=0$ and $\alpha_{ij}(r_c)=0$, $k(r_c)=0$, $B(r_c)=\frac{2 \partial_k\beta^k }{3f(r_c)\sqrt{f(r_c)}}$. By using the boundary conditions, these constants can be determined. Correspondingly, the three parameters $ b, C_B$ and $C_k$ are
\begin{equation}
C_B=0, C_k=-\frac{2 \partial_k \beta^k r_c^2}{3\sqrt{f(r_c)}}, b=0,
\end{equation}
and
\begin{equation}
C_B=\frac{2 \partial_k \beta^k}{3\sqrt{f(r_c)}}, C_k=\frac{4M\partial_k \beta^k}{3r_c^2f^{\frac{3}{2}}(r_c)}, b=0,
\end{equation}
and hence the corresponding first order stress tensor in (\ref{BoundaryST}) is explicitly obtained. For the first set of boundary conditions, the explicit first order stress tensor is
\begin{equation}
T_{vv}^{(1)}=-2 \partial_i \beta_i,~ T^{(1)}_{i j} = \frac{- 2r_+^3\sigma _{ij}}{r_c^3}+\frac{2(2M+r_c^4)}{3(2M-r_c^4)} \partial_k \beta_k \delta_{ij},
\label{t11}
\end{equation}
while the first order stress tensor in the second set of boundary conditions is
\begin{equation}
T_{vv}^{(1)}=0,~ T^{(1)}_{i j} = \frac{- 2r_+^3\sigma _{ij}}{r_c^3}.
\label{t12}
\end{equation}
Interestingly, we find that the spatial components of the two first order stress tensors in (\ref{t11}) and (\ref{t12}) have the same expressions. Naively, the above two stress tensor (\ref{t11}) and (\ref{t12}) in fact can be rewritten as the same form as
\begin{equation}
  T^{(1)}_{i j} = \frac{- 2r_+^3\sigma _{ij}}{r_c^3}+c_s^2T_{vv}^{(1)}{\delta_{ij}}, \label{Universality1}
\end{equation}
where $c_s^2=-\frac{2M+r_c^4}{3(2M-r_c^4)}$ is just related to the sound velocity of the holographic fluid. This point can be clearly seen in the followings. Note that, from (\ref{STBackground}) with $C=3$, the zero order pressure and energy density of dual fluid are $p_0=\frac{-4 M+2\left(3-3\sqrt{f\left(r_c\right)}\right)r_c^4}{r_c^4\sqrt{f\left(r_c\right)}}$, $\rho_0=2\left(3-3 \sqrt{f\left(r_c\right)} \right)$, and hence one will easily find that the sonic velocity satisfies,
\begin{eqnarray}
c_s^2=(\frac{\partial p_0}{\partial \rho_0})_{s}=\frac{(2 M +r_c^4)}{3( -2 M+r_c^4)}, \label{SoundV}
\end{eqnarray}
with the entropy density $s=\frac{r_+^3}{4 G r_c^3}$.

\subsection{The Second case: $h(r)=0$ and $B(r)=0$}
 A simple reason for choosing this case is that this gauge has not been considered before. We find that the function $F(r)$ will be nonzero in this case. Therefore, the residue functions need to be solved is $k(r)$ and $\alpha_{ij}(r)$. In this case, we can analytically solve the equations and the explicit results are
\begin{eqnarray}
\alpha_{ij}(r)&=&\alpha(r)\left\{(\partial_i \beta_j + \partial_j \beta_i )-\frac{2}{3} \delta_{ij}\partial_k \beta^k \right\} +(b-\frac{C_1}{3r}) \delta_{ij},\\
k(r)&=& C_{k2}+\frac{2r^3\partial_k \beta^k}{3r_c\sqrt{f(r_c)}}+\frac{C_1(r^4+2M)}{3f(r_c)r^2_cr},
\end{eqnarray}
Clearly, three parameters $b$, $C_1$ and $C_{k2}$ need to be fixed, which correspond to three boundary conditions. For the first set of boundary conditions under this gauge are $\alpha_{ij}(r_c)=0$, $k(r_c)=0$ and $F(r_c)=-\frac{2\partial_k \beta^k}{r_c \sqrt{f(r_c)}}$,%$T^{(1)}_{v v}(r_c)=0$%,
and the three parameters are
\begin{eqnarray}
C_1= \frac{-2r_c\partial_k \beta^k}{\sqrt{f(r_c)}},~~C_{k2}= \frac{8M\partial_k \beta^k}{3r^2_cf(r_c)\sqrt{f(r_c)}},~~ b=\frac{-2\partial_k \beta^k}{3\sqrt{f(r_c)}},
\end{eqnarray}
while the stress tensor is
\begin{equation}
T_{vv}^{(1)}=0,~~T^{(1)}_{ij} = \frac{- 2r_+^3\sigma _{ij}}{r_c^3}.\label{Case2ST1}
\end{equation}
Of course, we can also choose the other set of boundary conditions like $\alpha_{ij}(r_c)=0$, $k(r_c)=0$ and $F(r_c)=0$. In this case, the three parameters are
\begin{eqnarray}
C_1= 0,~~C_{k2}=-\frac{2 \partial_k \beta^k r_c^2}{3\sqrt{f(r_c)}},~~b=0.
\end{eqnarray}
 The corresponding stress tensor is
\begin{equation}
T_{vv}^{(1)}=-2 \partial_i \beta_i,~ T^{(1)}_{i j} = \frac{- 2r_+^3\sigma _{ij}}{r_c^3}+\frac{2(2M+r_c^4)}{3(2M-r_c^4)} \partial_k \beta_k \delta_{ij}. \label{Case2ST2}
\end{equation}
One sees that the two stress tensors in (\ref{Case2ST1}) and (\ref{Case2ST2}) can be also rewritten in the same form as Eq.(\ref{Universality1}). It should be noted that, the same form (\ref{Universality1}) is obtained in different boundary conditions under different gauges. From the two heuristic examples we conjecture that there may be an underlying universality behind (\ref{Universality1}). We investigate this possibility in the following section.

\section{An underlying relationship and the universality of the stress tensor}
From the above results with the same relation between $T^{(1)}_{v v}$ and $T^{(1)}_{ij}$ under different gauge and boundary conditions (\ref{Universality1}), one can conjecture that there may be an underlying universality in the stress tensor and some underlying relationships. Therefore, we investigate the possible underlying universality when all the gauge and boundary conditions are not fixed. For convenience and excluding the shear tensor in the stress tensor, we can first obtain the general solution of $\alpha_{ij}(r)$ from (\ref{A8})
\begin{eqnarray}
\alpha_{ij}(r)&=&\alpha(r)\left\{(\partial_i \beta_j + \partial_j \beta_i )-\frac{2}{3} \delta_{ij}\partial_k \beta^k \right\} +(\int^{r}\frac{F(s)}{3}ds+b) \delta_{ij}.
\end{eqnarray}
Then we define a useful quantity
\begin{equation}
P\equiv T_{xx}^{(1)}+2\eta\sigma_{xx}-c_s^2T_{vv}^{(1)}, \label{UsefulP}
\end{equation}
where $\eta=\frac{r_+^3}{r_c^3}$ is the shear viscosity of the holographic fluid at the finite cutoff surface. Substituting the results in (\ref{BoundaryST}) into (\ref{UsefulP}), we obtain
\begin{eqnarray}
P&=&2\partial_k\beta^k-4 \sqrt{f(r_c)} B(r_c)+\frac{(-2M+3r_c^4)F(r_c)}{3\sqrt{f(r_c)}r_c^3}+\frac{4 M-6r_c^4}{3 \sqrt{f(r_c)}r_c^3} h'(r_c)-\frac{\sqrt{f(r_c)}}{r_c}k'(r_c)\nonumber\\
&&+\frac{(2r_c^4+4M)(C-3\sqrt{f(r_c)})}{3 r_c^6f(r_c)}k(r_c)-2\left(C+\frac{2 M-3 r_c^4}{\sqrt{f\left(r_c\right)} r_c^4}\right)\alpha_{x x}\left(r_c\right)\nonumber\\
&&+\frac{4\left(2M-9r_c^4+C\sqrt{f(r_c)}r_c^4\right)}{3\sqrt{f(r_c)}r_c^4}h(r_c).
\end{eqnarray}
Note that, $P$ is zero in the above two cases, which implicates that $P$ may be related to $W_{AB}$ in the appendix~\ref{A}. Along this clue, we find that $P$ indeed is related to  $ W_{ii}$ and $W_{rr}$. In detail, we define a quantity $Q$ using the equations $ W_{ii}$ and $W_{rr}$
\begin{equation}
Q\equiv (\frac{r^4}{3r_c^2}f(r)W_{rr}-W_{xx})|_{r=r_c},
\end{equation}
which can be further rewritten as
\begin{equation}
Q=2 \sqrt{f(r_c)}\partial_k\beta^k-4f(r_c)B(r_c)+(-\frac{2 M}{3r_c^3}+r_c)
F(r_c)-8 h(r_c)+\frac{4 M-6r_c^4}{3r_c^3} h'(r_c)-\frac{f(r_c)}{r_c}k'(r_c).
\end{equation}
We find a relation between $P$ and $Q$
\begin{equation}
P=\frac{1}{\sqrt{f(r_c)}}Q+\frac{(2r_c^4+4M)(C-3\sqrt{f(r_c)})}{3r_c^6f(r_c)}k(r_c)-2\left(C+\frac{2 M-3 r_c^4}{\sqrt{f\left(r_c\right)} r_c^4}\right)\left(\alpha_{xx}\left(r_c\right)-\frac{2}{3}h(r_c)\right).
\end{equation}
This relation leads to an interesting result
\begin{eqnarray}
T_{xx}^{(1)}&=&-2\eta\sigma_{xx}+c_s^2T_{vv}^{(1)}+\frac{1}{\sqrt{f(r_c)}}(\frac{r^4}{3r_c^2}f(r)W_{rr}-W_{xx})|_{r=r_c}\nonumber\\
&+&\frac{(2r_c^4+4M)(C-3\sqrt{f(r_c)})}{3r_c^6f(r_c)}k(r_c)-2\left(C+\frac{2 M-3 r_c^4}{\sqrt{f\left(r_c\right)} r_c^4}\right)\left(\alpha_{x x}\left(r_c\right)-\frac{2}{3}h(r_c)\right), \label{Universality}
\end{eqnarray}
which explicitly expresses the universality for the first order stress tensor of holographic fluid at the finite cutoff surface. It should be noted that, the above universality can be treated as some kind of first order off-shell universality, because some of the first order on-shell conditions or equations $W_{AB}=0$ have not been used, i.e. $W_{rr}=W_{xx}=0$. Of course, the final stress tensor of holographic fluid should be on-shell, i.e. satisfying the first order on-shell equation $W_{AB}=0$, which is a direct result of the gravity field equation. Therefore, from (\ref{Universality}), the final universality in the first order stress tensor for holographic fluid is
\begin{equation}
T_{ij}^{(1)}=-2\eta\sigma_{ij}+c_s^2T_{vv}^{(1)}\delta_{ij}+\frac{(2r_c^4+4M)(C-3\sqrt{f(r_c)})}{3r_c^6f(r_c)}k(r_c)\delta_{ij}-2\left(C+\frac{2 M-3 r_c^4}{\sqrt{f\left(r_c\right)} r_c^4}\right)\left(\alpha_{ij}\left(r_c\right)-\frac{2}{3}h(r_c)\delta_{ij}\right), \label{UniversalityF}
\end{equation}
 One can check that the universal relation degenerates to the above two special cases after imposing the boundary conditions, i.e. $k(r_c)=0$ and $\alpha_{i i}\left(r_c\right)=\frac{2}{3}h(r_c)$. The universality clearly displays the relation between the first order components of the stress tensor $T_{ij}^{(1)},~T_{vv}^{(1)}$, which is free of gauge and boundary condition choices.

\section{Conclusion and discussion}
We find a universality in the fluid/gravity correspondence. That is, the first order stress tensors of the holographic fluids located at the boundary are essentially identical for different gauge and boundary conditions of the metric perturbations in the bulk, through a comprehensive investigation on the holographic fluids at the finite cutoff surface in the perturbated Schwarschild-AdS black brane spacetime.

Different from the previous works, we derive full correction terms in the metric perturbation, and obtain the corresponding general stress tensor. A simple reason of taking the full correction terms into account is that we can choose more boundary conditions under different gauge conditions at the finite cutoff surface. Moreover, it is more convenient to explicitly find  the universality in the first order stress tensor. In two heuristic examples, we  explicitly show the stress tensors. An interesting result is that these two different stress tensors in fact can be rewritten as the same form, which implicates that there is a universality in the first order stress tensor of holographic fluid at the finite cutoff surface. By investigating the general first order stress tensor, and making comparison with the tensor $W_{AB}$ related to the first order perturbated equations, we find the underlying relationship between the first order stress tensor and $W_{AB}$, and hence the underlying universality in the first order stress tensor of holographic fluid emerges.
After using the the first order perturbative equation $W_{AB}=0$, we derive the universality (\ref{UniversalityF}) in the first order stress tensor, which may be related to the realistic transport coefficients like the shear and bulk viscosities~\footnote{The work of using this universality to extract the true transport coefficients is in process.}.

It should be emphasized that this universality in the first order stress tensor in (\ref{Universality}) or (\ref{UniversalityF}) is independent on the different gauge and boundary conditions. One of our key observation is to introduce the sonic velocity in this universal relation. The second interesting result is that the universality in (\ref{Universality}) can be treated as some underlying first order off-shell relation, since the on-shell condition, i.e. the first order perturbative equation $W_{AB}=0$, is not invoked. Whether or not there is some underlying physics in this first order off-shell relation is an interesting and open topic. Five gauge freedoms exist in the first order full correction terms for the metric perturbations. Therefore, usually there are five constraint equations in $W_{AB}=0$. In our case, four of them correspond to the conservation of the zero stress tensor of the holographic fluid at the cutoff surface, and the fifth constraint equation automatically satisfies when one turns on the on-shell equations $W_{AB}=0$. We find a second type constraint equation in the $W_{AB}$, which is similar to the universality in the first order stress tensor. More details about these constraint equations can be found in the appendix ~\ref{B}. Therefore, some underlying physics may exist in this first order off-shell constraint equation, and whether or not there is some relationship between the universality and this constraint equation deserves to study further. Investigations on more different background solutions other than the Schwarschild-AdS may present some insights on these questions in the future works.

\newpage

\appendix
\section{The tensor components of $W_{AB }$ and $S_{AB }$}
\label{A} The tensor components of $W_{AB } = (\text{effect
from correction}) - S_{AB}$ are
\begin{eqnarray}
W_{vv} &=& -\frac{8 r^2 f(r)}{r_c^2 f(r_c)} h(r) - \frac{4(2M^2-Mr^4+r^8)f(r)}{r^6 r_c^2 f(r_c)} B(r) -\frac{(2 M+r^4) f(r)}{r r_c^2 f(r_c)} \left(2h'(r) - \frac{F(r)}{2} \right)  \\ \nonumber
& & +\frac{f(r)}{2 r} k'(r) - \frac{f^2(r) (2M + r^4)}{2rr_c^2f(r_c)}B'(r) -\frac{1}{2} f(r) k''(r) -S_{vv}^{(1)},\nonumber\\
W_{vi} &=& \frac{3 f(r)}{2 r} j_i'(r) -\frac{1}{2} f(r) j_i''(r)-S_{vi}^{(1)}(r)~,\\
W_{vr} &=& \frac{8}{r_c \sqrt{f(r_c)}}h(r) +\frac{4 (2M^2-Mr^4+r^8)}{r^8r_c\sqrt{f(r_c)}} B(r) +\frac{(2 M +r^4)}{r^3 r_c \sqrt{f(r_c)}} \left(2h'(r) - \frac{F(r)}{2} \right) \\ \nonumber
& & -\frac{r_c \sqrt{f(r_c)}}{2 r^3}k'(r) + \frac{(2M+r^4)f(r)}{2r^3\sqrt{f(r_c)r_c}}B'(r) +\frac{r_c \sqrt{f(r_c)}}{2 r^2}k''(r) -S_{vr}^{(1)} \\
W_{ri} &=& -\frac{3 r_c \sqrt{f(r_c)}}{2 r^3}j_i'(r) +\frac{r_c \sqrt{f(r_c)}}{2 r^2}j_i''(r) -S_{ri}^{(1)}\\
W_{rr} &=& \frac{12 M}{r^6}B(r) +\frac{5}{r} h'(r) +\frac{3f(r)}{2r}B'(r) + h''(r)-\frac{1}{r}F(r) -\frac{1}{2}F'(r) -S_{rr}^{(1)}\\
W_{ii} &=& \frac{8 r^2}{r_c^2} h(r) +\frac{4f(r)(M+r^4)}{r^2r_c^2}B(r) +\frac{f(r)r^3}{r_c^2}h'(r) +\frac{8(-M+r^4)}{3rr_c^2}h'(r) +\frac{f(r_c)}{r}k'(r) + \frac{f^2(r) r^3}{2 r_c^2}B'(r)\nonumber\\
& &+\frac{r^4 f(r)}{3r_c^2}h''(r)+\frac{\left(2 M -5 r^4\right) \alpha _{i i}'(r)}{2 r r_c^2}-\frac{1}{2 r_c^2} r^4 f(r) \alpha _{i i}''(r)-\frac{r^3f(r)}{2r_c^2}F(r)-S_{ii}^{(1)} \label{A6}\\
W_{ij} &=& \frac{\left(2 M -5 r^4\right) \alpha _{i j}'(r)}{2 r r_c^2}-\frac{1}{2r_c^2} r^4 f(r) \alpha _{i j}''(r)-S_{ij}^{(1)},~(i\neq j)\\
W_{ij} &-& \dfrac{1}{3}\delta_{ij}\left(\sum_k W_{kk}\right) = \frac{\left(2 M -5 r^4\right) \left(\alpha _{i j}'(r)- \delta_{ij} \frac{1}{3} F(r)\right)}{2 r r_c^2} -\frac{1}{2r_c^2} r^4 f(r) \left(\alpha _{i j}'(r) -\delta_{ij} \frac{1}{3} F(r)\right)' \nonumber\\
& & \quad\quad\quad \quad\quad\quad\quad\quad+\dfrac{1}{3}\delta_{ij}(\delta^{kl}S_{kl}^{(1)}) -S_{ij}^{(1)} \label{A8}
\end{eqnarray}
where $F(r)=\sum_i\alpha _{i i}'(r)$, and the first order source terms are
\begin{eqnarray}
S_{vv}^{(1)}(r)&=&-\frac{3\partial _vM}{r^3 r_c \sqrt{f\left(r_c\right)}}-\frac{\left(2 M +r^4\right) \partial _i\beta _i}{r^3 r_c \sqrt{f\left(r_c\right)}},\\
S_{vi}^{(1)}(r)&=&\frac{\left(-2 M +3 r^4+2 r_c^4\right) \partial _iM}{2 r^3 r_c^5 f\left(r_c\right){}^{3/2}}+\frac{\left(2 M +3 r^4\right) \partial _v\beta _i}{2 r^3 r_c \sqrt{f\left(r_c\right)}},\\
S_{vr}^{(1)}(r)&=&\frac{\partial _i\beta _i}{r},\\
S_{ri}^{(1)}(r)&=&-\frac{3 \partial _v\beta _i}{2 r}-\frac{3 \partial _iM}{2 r r_c^4 f\left(r_c\right)},\\
S_{rr}^{(1)}(r) &=&0, \\
S_{ij}^{(1)}(r) &=&\left(\delta _{i j}\partial _k\beta _k+3\partial _{(i}\beta _{j)}\right)\frac{r\sqrt{f(r_c)}}{r_c}.
\end{eqnarray}

\section{The constraint equations}
\label{B}
From the first order perturbative equations $W_{AB}=0$, we can find that the cutoff effect has been incorporated through their dependence on $r_c$, while there are five constraint equations
\begin{eqnarray}
 &&W_{vv} + \dfrac{r^2 f(r)}{r_c\sqrt{f(r_c)}}W_{vr} =0 ~\Rightarrow~ S_{vv}^{(1)} + \dfrac{r^2 f(r)}{r_c\sqrt{f(r_c)}}S_{vr}^{(1)} = 0, \notag \\
 &&W_{vi} + \dfrac{r^2 f(r)}{r_c\sqrt{f(r_c)}} W_{ri} =0 ~\Rightarrow~ S_{vi}^{(1)} +\dfrac{r^2 f(r)}{r_c\sqrt{f(r_c)}}S_{ri}^{(1)} = 0, \label{constraint}
 \end{eqnarray}
after using the first order source terms in the appendix~\ref{A}, one can further rewrite these constraint equations~(\ref{constraint}) as
\begin{eqnarray}
&&3 \partial _vM+4 M \partial _i\beta _i=0,\\\nonumber
&& \partial _iM+4 M \partial _v\beta _i=\frac{- 4 M \partial _i M }{r_c^4 f\left(r_c\right)}~~,
\end{eqnarray}
which are just the conservation equations of the zeroth order stress-energy tensor~\cite{Bhattacharyya:2008jc, Hur:2008tq,Hu:2010sn, Hu:2011ze}. On the other hand, after some physical arguments or direct calculations, these four constraint equations are also contained in the five constraint equations $W_{AB}n^A = 0$, where $n^A$ is the normal vector of the cutoff surface, i.e. $W_{Av}n^A=0$ and $W_{Ai}n^A=0$. For the fifth constraint equation $W_{Ar}n^A=0$, it is a equation related to the $W_{vr}$ and $W_{rr}$, which can be automatically consistent with the first order on-shell conditions or equations $W_{AB}=0$. Note that, an interesting discovery is that we have further found another constraint equation
\begin{equation}
\frac{r_c}{2 \sqrt{f(r_c)} r} W_{ii}^{'} - W_{vr} + \frac{2M-r^4}{6r \sqrt{f(r_c)}r_c} W'_{rr} + \frac{2M-5r^4}{3r^2\sqrt{f(r_c)} r_c} W_{rr}= 0,  \label{constraint2}
\end{equation}
where the index $(ii)$ does not represent the sum here. However, it should be emphasized that this constraint equation (\ref{constraint2}) is in fact different from the above four constraint equations (\ref{constraint}). Since the above four constraint equations (\ref{constraint}) can be considered as the first order on-shell constraint equations, while (\ref{constraint2}) is the first order off-shell since the first order on-shell condition $W_{AB}=0$ has not been used.

\acknowledgments

Y.P Hu thanks a lot for the discussions with Profs.Rong-Gen Cai, Li-Ming Cao, Koenraad Schalm, Jan Zaanen, Hai-Qing Zhang and Drs. Saso Grozdanov, Song He, Yun-Long Zhang, Debarghya Banerjee. This work is supported by the National Natural Science Foundation of China (NSFC) (Grant Nos.11575083, 11565017, 11275128, 11105004 and 11075106), the Fundamental Research Funds for the Central Universities (Grant No. NS2015073), and the Open Project Program of State Key Laboratory of Theoretical Physics, Institute of Theoretical Physics, Chinese Academy of Sciences, China (Grant  No. Y5KF161CJ1). In addition, Y.P Hu thanks a lot for the support from the Sino-Dutch scholarship program under the China Scholarship Council (CSC).

%\paragraph{Note added.} This is also a good position for notes added after the paper has been written.



\begin{thebibliography}{99}

%\cite{Zaanen:2015oix}
\bibitem{Zaanen:2015oix}
  J.~Zaanen, Y.~W.~Sun, Y.~Liu and K.~Schalm,
  ``Holographic Duality in Condensed Matter Physics;''
  %%CITATION = INSPIRE-1384852;%%
  %5 citations counted in INSPIRE as of 21 Mar 2017
%\cite{Cubrovic:2009ye}
%\bibitem{Cubrovic:2009ye}
  M.~Cubrovic, J.~Zaanen and K.~Schalm,
  %``String Theory, Quantum Phase Transitions and the Emergent Fermi-Liquid,''
  Science {\bf 325}, 439 (2009)
  %doi:10.1126/science.1174962
  [arXiv:0904.1993 [hep-th]].
  %%CITATION = doi:10.1126/science.1174962;%%
  %318 citations counted in INSPIRE as of 21 Mar 2017


%\cite{Ammon:2015wua}
\bibitem{Ammon:2015wua}
  M.~Ammon and J.~Erdmenger,
  ``Gauge/gravity duality : Foundations and applications.''
  %%CITATION = INSPIRE-1376202;%%
  %3 citations counted in INSPIRE as of 08 May 2017

%\cite{CasalderreySolana:2011us}
\bibitem{CasalderreySolana:2011us}
  J.~Casalderrey-Solana, H.~Liu, D.~Mateos, K.~Rajagopal and U.~A.~Wiedemann,
  %``Gauge/String Duality, Hot QCD and Heavy Ion Collisions,''
  book:Gauge/String Duality, Hot QCD and Heavy Ion Collisions. Cambridge, UK: Cambridge University Press, 2014
  %doi:10.1017/CBO9781139136747
  [arXiv:1101.0618 [hep-th]].
  %%CITATION = doi:10.1017/CBO9781139136747;%%
  %454 citations counted in INSPIRE as of 08 May 2017




%%%%%%%%%%%%%%%%%%%%%%%%%%%%%%%%%%%%%%%%%%%%%%%%%%%%%%%%%%%%%%%%%%%%%%%%%%%%%%%%%%AdS/CFT correspondence
%\begin{thebibliography}{99}
\baselineskip 12pt
%\cite{Maldacena:1997re}
\bibitem{Maldacena:1997re}
  J.~M.~Maldacena,
  %``The large N limit of superconformal field theories and supergravity,''
  Adv.\ Theor.\ Math.\ Phys.\  {\bf 2}, 231 (1998)
  [Int.\ J.\ Theor.\ Phys.\  {\bf 38}, 1113 (1999)]
  [arXiv:hep-th/9711200].
  %%CITATION = IJTPB,38,1113;%%
%\cite{Gubser:1998bc}
\bibitem{Gubser:1998bc}
  S.~S.~Gubser, I.~R.~Klebanov and A.~M.~Polyakov,
  %``Gauge theory correlators from non-critical string theory,''
  Phys.\ Lett.\  B {\bf 428}, 105 (1998)
  [arXiv:hep-th/9802109].
  %%CITATION = PHLTA,B428,105;%%
%\cite{Witten:1998qj}
\bibitem{Witten:1998qj}
  E.~Witten,
  %``Anti-de Sitter space and holography,''
  Adv.\ Theor.\ Math.\ Phys.\  {\bf 2}, 253 (1998)
  [arXiv:hep-th/9802150].
  %%CITATION = 00203,2,253;%%
%\cite{Aharony:1999ti}
\bibitem{Aharony:1999ti}
  O.~Aharony, S.~S.~Gubser, J.~M.~Maldacena, H.~Ooguri and Y.~Oz,
  %``Large N field theories, string theory and gravity,''
  Phys.\ Rept.\  {\bf 323}, 183 (2000)
  [arXiv:hep-th/9905111].
  %%CITATION = PRPLC,323,183;%%
%\cite{Herzog:2009xv}
\bibitem{Herzog:2009xv}
  C.~P.~Herzog,
  %``Lectures on Holographic Superfluidity and Superconductivity,''
  J.\ Phys.\ A  {\bf 42}, 343001 (2009)
  [arXiv:0904.1975 [hep-th]];
  %%CITATION = JPAGB,A42,343001;%%
%\cite{Brihaye:2010mr}
%\bibitem{Brihaye:2010mr}
  Y.~Brihaye and B.~Hartmann,
  %``Holographic Superconductors in 3+1 dimensions away from the probe limit,''
  Phys.\ Rev.\  D {\bf 81}, 126008 (2010)
  [arXiv:1003.5130 [hep-th]].
  %%CITATION = PHRVA,D81,126008;%%


%%%%%%%%%%%%%%%%%%%%%%%%%%%%%%%%%%%%%%%%%%%%%%%%%%%%%%%%%%%%%%%%%%%%%%%%%%%%%%%%%shear viscosity to entropy density ration via AdS/CFT correspondence

%\cite{Policastro:2001yc}
\bibitem{Policastro:2001yc}
  G.~Policastro, D.~T.~Son and A.~O.~Starinets,
  %``The shear viscosity of strongly coupled N = 4 supersymmetric Yang-Mills
  %plasma,''
  Phys.\ Rev.\ Lett.\  {\bf 87}, 081601 (2001)
  [arXiv:hep-th/0104066].
  %%CITATION = PRLTA,87,081601;%%

%\cite{Buchel:2003tz}
\bibitem{Buchel:2003tz}
  A.~Buchel and J.~T.~Liu,
  %``Universality of the shear viscosity in supergravity,''
  Phys.\ Rev.\ Lett.\  {\bf 93}, 090602 (2004)
  [arXiv:hep-th/0311175].
  %%CITATION = PRLTA,93,090602;%%

%%%%%%%%%%%%%%%%%%%%%%%%%%%%%%%%%%%%%%%%%%%%%%%%%%KSS bound

%\cite{Kovtun:2003wp}
\bibitem{Kovtun:2003wp}
  P.~Kovtun, D.~T.~Son and A.~O.~Starinets,
  %``Holography and hydrodynamics: Diffusion on stretched horizons,''
  JHEP {\bf 0310}, 064 (2003)
  [arXiv:hep-th/0309213].
  %%CITATION = JHEPA,0310,064;%%
  %\cite{Buchel:2003tz}

%\cite{Kovtun:2004de}
\bibitem{Kovtun:2004de}
  P.~Kovtun, D.~T.~Son and A.~O.~Starinets,
  %``Viscosity in strongly interacting quantum field theories from black hole
  %physics,''
  Phys.\ Rev.\ Lett.\  {\bf 94}, 111601 (2005)
  [arXiv:hep-th/0405231].
  %%CITATION = PRLTA,94,111601;%%


%%%%%%%%%%%%%%%%%%%%%%%%%%%%%%%%%%%%%%%%%%%%%%%%%%%%%%%%%%%%%%%%%shear viscosity to entropy density ration in the experiment data
%\cite{Song:2010mg}
\bibitem{Song:2010mg}
  H.~Song, S.~A.~Bass, U.~Heinz, T.~Hirano and C.~Shen,
  %``200 A GeV Au+Au collisions serve a nearly perfect quark-gluon liquid,''
  Phys.\ Rev.\ Lett.\  {\bf 106}, 192301 (2011)
  Erratum: [Phys.\ Rev.\ Lett.\  {\bf 109}, 139904 (2012)]
  %doi:10.1103/PhysRevLett.106.192301, 10.1103/PhysRevLett.109.139904
  [arXiv:1011.2783 [nucl-th]].
  %%CITATION = doi:10.1103/PhysRevLett.106.192301, 10.1103/PhysRevLett.109.139904;%%
  %279 citations counted in INSPIRE as of 08 May 2017



%%%%%%%%%%%%%%%%%%%%%%%%%%%%%%%%%%%%%%%%%%%%%%%%%% fluid-gravity correspondence
%\cite{Bhattacharyya:2008jc,Rangamani:2009xk}

%\cite{Bhattacharyya:2008jc}
\bibitem{Bhattacharyya:2008jc}
  S.~Bhattacharyya, V.~E.~Hubeny, S.~Minwalla and M.~Rangamani,
  %``Nonlinear Fluid Dynamics from Gravity,''
  JHEP {\bf 0802}, 045 (2008)
  [arXiv:0712.2456 [hep-th]];
  %%CITATION = JHEPA,0802,045;%%


%\cite{Rangamani:2009xk}
\bibitem{Rangamani:2009xk}
  M.~Rangamani,
  %``Gravity and Hydrodynamics: Lectures on the fluid-gravity correspondence,''
  Class.\ Quant.\ Grav.\  {\bf 26}, 224003 (2009)
  %doi:10.1088/0264-9381/26/22/224003
  [arXiv:0905.4352 [hep-th]].
  %%CITATION = doi:10.1088/0264-9381/26/22/224003;%%
  %189 citations counted in INSPIRE as of 19 Feb 2017









%%%%%%%%%%%%%%%%%%%%%%%%%%%%%%%%%%%%%%%%Gravity/Hydrodynamics correspondence applied
%%%%%%%%%%%%%%%%%%%%%%%%%%Hur:2008tq,Erdmenger:2008rm,Banerjee:2008th,Tan:2009yg
%Torabian:2009qk,Hu:2010sn,Anninos:2008sj,Cvetic:2001bk
%\cite{Hur:2008tq}
\bibitem{Hur:2008tq}
  J.~Hur, K.~K.~Kim and S.~J.~Sin,
  %``Hydrodynamics with conserved current from the gravity dual,''
  JHEP {\bf 0903}, 036 (2009)
  [arXiv:0809.4541 [hep-th]].
  %%CITATION = JHEPA,0903,036;%%

%\cite{Erdmenger:2008rm}
\bibitem{Erdmenger:2008rm}
  J.~Erdmenger, M.~Haack, M.~Kaminski and A.~Yarom,
  %``Fluid dynamics of R-charged black holes,''
  JHEP {\bf 0901}, 055 (2009)
  [arXiv:0809.2488 [hep-th]].
  %%CITATION = JHEPA,0901,055;%%

%\cite{Banerjee:2008th}
\bibitem{Banerjee:2008th}
  N.~Banerjee, J.~Bhattacharya, S.~Bhattacharyya, S.~Dutta, R.~Loganayagam and P.~Surowka,
  %``Hydrodynamics from charged black branes,''
 JHEP {\bf 1101}, 094 (2011)  [arXiv:0809.2596 [hep-th]].
  %%CITATION = ARXIV:0809.2596;%%

%%%%%%%%%%%%%%%%%%%%%%%%%%%%%%%%%%%%%%%%%%%%%%%%%%%%%%%%%%%%%Chern-Simons terms
%\cite{Son:2009tf}
\bibitem{Son:2009tf}
  D.~T.~Son and P.~Surowka,
  %``Hydrodynamics with Triangle Anomalies,''
  Phys.\ Rev.\ Lett.\  {\bf 103}, 191601 (2009)
  [arXiv:0906.5044 [hep-th]].
  %%CITATION = PRLTA,103,191601;%%


%\cite{Tan:2009yg}
\bibitem{Tan:2009yg}
  H.~S.~Tan,
  %``Born-Infeld Hydrodynamics via Gauge/Gravity Duality,''
  JHEP {\bf 0904}, 131 (2009)
  [arXiv:0903.3424 [hep-th]].
  %%CITATION = JHEPA,0904,131;%%



%\cite{Torabian:2009qk}
\bibitem{Torabian:2009qk}
  M.~Torabian and H.~U.~Yee,
  %``Holographic nonlinear hydrodynamics from AdS/CFT with multiple/non-Abelian
  %symmetries,''
  JHEP {\bf 0908}, 020 (2009)
  [arXiv:0903.4894 [hep-th]].
  %%CITATION = JHEPA,0908,020;%%




%\cite{Hu:2010sn}
\bibitem{Hu:2010sn}
  Y.~P.~Hu, H.~F.~Li and Z.~Y.~Nie,
  %``The first order hydrodynamics via AdS/CFT correspondence in the
  %Gauss-Bonnet gravity,''
  JHEP {\bf 1101}, 123 (2011)
  [arXiv:1012.0174 [hep-th]];
  %%CITATION = JHEPA,1101,123;%%
%\cite{He:2011hw}
%\bibitem{He:2011hw}
  S.~He, Y.~P.~Hu and J.~H.~Zhang,
  %``Hydrodynamics of a 5D Einstein-dilaton black hole solution and the corresponding BPS state,''
  JHEP {\bf 1112}, 078 (2011)
  %doi:10.1007/JHEP12(2011)078
  [arXiv:1111.1374 [hep-th]].
  %%CITATION = doi:10.1007/JHEP12(2011)078;%%
  %7 citations counted in INSPIRE as of 27 Jun 2017


%\cite{Hu:2011ze}
\bibitem{Hu:2011ze}
  Y.~-P.~Hu, P.~Sun and J.~-H.~Zhang,
  %``Hydrodynamics with conserved current via AdS/CFT correspondence in the
  %Maxwell-Gauss-Bonnet gravity,''
  Phys.\ Rev.\  D {\bf 83}, 126003 (2011)
  [arXiv:1103.3773 [hep-th]];
  %%CITATION = PHRVA,D83,126003;%%
%\cite{Hu:2011qa}
%\bibitem{Hu:2011qa}
  Y.~-P.~Hu and C.~-Y.~Park,
  %``Chern-Simons effect on the dual hydrodynamics in the Maxwell-Gauss-Bonnet gravity,''
  Phys.\ Lett.\ B {\bf 714}, 324 (2012)
  [arXiv:1112.4227 [hep-th]];
  %%CITATION = ARXIV:1112.4227;%%
%\cite{Hu:2013dza}
%\bibitem{Hu:2013dza}
  Y.~P.~Hu and J.~H.~Zhang,
  %``Gravity/Fluid Correspondence and Its Application on Bulk Gravity with $U(1)$ Gauge Field,''
  Adv.\ High Energy Phys.\  {\bf 2014}, 483814 (2014)
  [arXiv:1311.3974 [hep-th]].
  %%CITATION = ARXIV:1311.3974;%%
  %1 citations counted in INSPIRE as of 09 Aug 2014



%\cite{Kalaydzhyan:2011vx}
\bibitem{Kalaydzhyan:2011vx}
  T.~Kalaydzhyan and I.~Kirsch,
  %``Fluid/gravity model for the chiral magnetic effect,''
  Phys.\ Rev.\ Lett.\  {\bf 106}, 211601 (2011)
  [arXiv:1102.4334 [hep-th]].
  %%CITATION = PRLTA,106,211601;%%


%\cite{Amado:2011zx}
\bibitem{Amado:2011zx}
  I.~Amado, K.~Landsteiner and F.~Pena-Benitez,
  %``Anomalous transport coefficients from Kubo formulas in Holography,''
  JHEP {\bf 1105}, 081 (2011)
  [arXiv:1102.4577 [hep-th]];
  %%CITATION = ARXIV:1102.4577;%%
%\cite{Landsteiner:2012dm}
%\cite{Landsteiner:2011tf}
%\bibitem{Landsteiner:2011tf}
  K.~Landsteiner, E.~Megias, L.~Melgar and F.~Pena-Benitez,
  %``Gravitational Anomaly and Hydrodynamics,''
  J.\ Phys.\ Conf.\ Ser.\  {\bf 343}, 012073 (2012)
  [arXiv:1111.2823 [hep-th]];
  %%CITATION = ARXIV:1111.2823;%%
%\bibitem{Landsteiner:2012dm}
  K.~Landsteiner and L.~Melgar,
  %``Holographic Flow of Anomalous Transport Coefficients,''
  arXiv:1206.4440 [hep-th].
  %%CITATION = ARXIV:1206.4440;%%
%\cite{Banerjee:2012iz}





%%%%%%%%%%%%%%%%%%%%%%%%%%%%%%%%%%%%%%%%%%%%%%%%%%%%%%%%%%%%%%%%%%%Cutoff relations between Navier-Stokes and Einstein equations

%%Bredberg:2010ky,Bredberg:2011jq,Compere:2011dx
%\cite{Bredberg:2010ky}
\bibitem{Bredberg:2010ky}
  I.~Bredberg, C.~Keeler, V.~Lysov and A.~Strominger,
  %``Wilsonian Approach to Fluid/Gravity Duality,''
  JHEP {\bf 1103}, 141 (2011)
  arXiv:1006.1902 [hep-th].
  %%CITATION = JHEPA,1103,141;%%





%%%%%%%%%%%%%%%%%%%%%%%%%%%%%%%%%%%%%%%%%%%%%%%%%%%%%%%%%%%%%%%%%%%Cutoff relations between Navier-Stokes and Einstein equations

%\cite{Bredberg:2011jq}
\bibitem{Bredberg:2011jq}
  I.~Bredberg, C.~Keeler, V.~Lysov and A.~Strominger,
  %``From Navier-Stokes To Einstein,''
  JHEP {\bf 1207}, 146 (2012)
  %doi:10.1007/JHEP07(2012)146
  [arXiv:1101.2451 [hep-th]].
  %%CITATION = doi:10.1007/JHEP07(2012)146;%%
  %121 citations counted in INSPIRE as of 16 May 2017


%\cite{Compere:2011dx}
\bibitem{Compere:2011dx}
  G.~Compere, P.~McFadden, K.~Skenderis and M.~Taylor,
  %``The Holographic fluid dual to vacuum Einstein gravity,''
  JHEP {\bf 1107}, 050 (2011)
  arXiv:1103.3022 [hep-th].
  %%CITATION = JHEPA,1107,050;%%


%%Cai:2011xv,Niu:2011gu,Niu:2011gu,Compere:2012mt,Eling:2012ni
%\cite{Cai:2011xv}
\bibitem{Cai:2011xv}
  R.~-G.~Cai, L.~Li and Y.~-L.~Zhang,
  %``Non-Relativistic Fluid Dual to Asymptotically AdS Gravity at Finite Cutoff Surface,''
  JHEP {\bf 1107}, 027 (2011)  [arXiv:1104.3281 [hep-th]];  %%CITATION = ARXIV:1104.3281;%%
%\cite{Cai:2012vr}
%\bibitem{Cai:2012vr}
  R.~-G.~Cai, L.~Li, Z.~-Y.~Nie and Y.~-L.~Zhang,
  %``Holographic Forced Fluid Dynamics in Non-relativistic Limit,''
 Nucl.\ Phys.\ B {\bf 864}, 260 (2012)  [arXiv:1202.4091 [hep-th]].  %%CITATION = ARXIV:1202.4091;%%


%\cite{Niu:2011gu,Compere:2012mt,Eling:2012ni}
\bibitem{Niu:2011gu}
  C.~Niu, Y.~Tian, X.-N.~Wu and Y.~Ling,
  %``Incompressible Navier-Stokes Equation from Einstein-Maxwell and Gauss-Bonnet-Maxwell Theories,''
  Phys.\ Lett.\ B {\bf 711}, 411 (2012) [arXiv:1107.1430 [hep-th]];  %%CITATION = ARXIV:1107.1430;%%
%\cite{Huang:2011he}
%\bibitem{Huang:2011he}
  T.-Z.~Huang, Y.~Ling, W.-J.~Pan, Y.~Tian and X.-N.~Wu,
  %``From Petrov-Einstein to Navier-Stokes in Spatially Curved Spacetime,''
  JHEP {\bf 1110}, 079 (2011) [arXiv:1107.1464 [gr-qc]];
  C.-Y. Zhang, Y. Ling, C. Niu, Y. Tian and X.-N. Wu, Phys. Rev. D {\bf 86}, 084043 (2012).  %%CITATION = ARXIV:1107.1464;%%



%\cite{Compere:2012mt,Eling:2012ni}
\bibitem{Compere:2012mt}
  G.~Compere, P.~McFadden, K.~Skenderis and M.~Taylor,
  %``The relativistic fluid dual to vacuum Einstein gravity,''
  JHEP {\bf 1203}, 076 (2012)  [arXiv:1201.2678 [hep-th]].
  %%CITATION = ARXIV:1201.2678;%%

%\cite{Eling:2012ni}
\bibitem{Eling:2012ni}
  C.~Eling, A.~Meyer and Y.~Oz,
  %``The Relativistic Rindler Hydrodynamics,''
  JHEP {\bf 1205}, 116 (2012)  [arXiv:1201.2705 [hep-th]].
  %%CITATION = ARXIV:1201.2705;%%

%%%%%%%%%%%%%%%%%%%%%%%%%%%%%%%%%%%%%%%%%%%%%%%%%%%%%%%%%%%%%%%%%%%%%%%%%%%%%%%%% radial coordinate as the energy scale

%\cite{Balasubramanian:1998de}
\bibitem{Balasubramanian:1998de}
  V.~Balasubramanian, P.~Kraus, A.~E.~Lawrence and S.~P.~Trivedi,
  %``Holographic probes of anti-de Sitter space-times,''
  Phys.\ Rev.\ D {\bf 59}, 104021 (1999)
  [hep-th/9808017].
  %%CITATION = HEP-TH/9808017;%%

%\cite{Susskind:1998dq}
\bibitem{Susskind:1998dq}
  L.~Susskind and E.~Witten,
  %``The Holographic bound in anti-de Sitter space,''
  hep-th/9805114.  %%CITATION = HEP-TH/9805114;%%


%\cite{Akhmedov:1998vf}
\bibitem{Akhmedov:1998vf}
  E.~T.~Akhmedov,
  %``A Remark on the AdS / CFT correspondence and the renormalization group flow,''
  Phys.\ Lett.\ B {\bf 442}, 152 (1998)
  [hep-th/9806217].
  %%CITATION = HEP-TH/9806217;%%

%\cite{de Boer:1999xf}
\bibitem{de Boer:1999xf}
  J.~de Boer, E.~P.~Verlinde and H.~L.~Verlinde,
  %``On the holographic renormalization group,''
  JHEP {\bf 0008}, 003 (2000)  [hep-th/9912012].  %%CITATION = HEP-TH/9912012;%%


%\cite{Kuperstein:2011fn,}
\bibitem{Kuperstein:2011fn}
  S.~Kuperstein and A.~Mukhopadhyay,
  %``The unconditional RG flow of the relativistic holographic fluid,''
  JHEP {\bf 1111}, 130 (2011)  [arXiv:1105.4530 [hep-th]];
%%CITATION = ARXIV:1105.4530;%%





%%%%%%%%%%%%%%%%%%%%%%%%%%%%%%%%%%%%%%%%%%%%%%%%%%%holographic renormalization grpups flow
%%Heemskerk:2010hk,Faulkner:2010jy,Iqbal:2008by,Sin:2011yh
%\cite{Heemskerk:2010hk}
\bibitem{Heemskerk:2010hk}
  I.~Heemskerk and J.~Polchinski,
  %``Holographic and Wilsonian Renormalization Groups,''
  JHEP {\bf 1106}, 031 (2011)  [arXiv:1010.1264 [hep-th]].  %%CITATION = ARXIV:1010.1264;%%

%\cite{Faulkner:2010jy}
\bibitem{Faulkner:2010jy}
  T.~Faulkner, H.~Liu and M.~Rangamani,
  %``Integrating out geometry: Holographic Wilsonian RG and the membrane paradigm,''
  JHEP {\bf 1108}, 051 (2011)  [arXiv:1010.4036 [hep-th]].  %%CITATION = ARXIV:1010.4036;%%

%\cite{Iqbal:2008by}
\bibitem{Iqbal:2008by}
  N.~Iqbal and H.~Liu,
  %``Universality of the hydrodynamic limit in AdS/CFT and the membrane paradigm,''
  Phys.\ Rev.\ D {\bf 79}, 025023 (2009)  [arXiv:0809.3808 [hep-th]].  %%CITATION = ARXIV:0809.3808;%%



%\cite{Sin:2011yh}
\bibitem{Sin:2011yh}
  S.~-J.~Sin and Y.~Zhou,
  %``Holographic Wilsonian RG Flow and Sliding Membrane Paradigm,''
  JHEP {\bf 1105}, 030 (2011)
  [arXiv:1102.4477 [hep-th]];  %%CITATION = ARXIV:1102.4477;%%
%\cite{Matsuo:2011fk}
%\bibitem{Matsuo:2011fk}
  Y.~Matsuo, S.~-J.~Sin and Y.~Zhou,
  %``Mixed RG Flows and Hydrodynamics at Finite Holographic Screen,''
  JHEP {\bf 1201}, 130 (2012)
  [arXiv:1109.2698 [hep-th]];  %%CITATION = ARXIV:1109.2698;%%
%\cite{Matsuo:2012bv}
%\bibitem{Matsuo:2012bv}
  Y.~Matsuo, S.~-J.~Sin and Y.~Zhou,
  %``Holographic RG Flow and Sound Modes of sQGP,''
  JHEP {\bf 1207}, 050 (2012)
  [arXiv:1204.6627 [hep-th]].  %%CITATION = ARXIV:1204.6627;%%




%%%%%%%%%%%%%%%%%%%%%%%%%%%%%%%%%%%%%%%%%%%%%%%%%%%%%%%%%%%%%%%%%%%%%%at finite cutoff surface and bulk viscosity

%\cite{Brattan:2011my}
\bibitem{Brattan:2011my}
  D.~Brattan, J.~Camps, R.~Loganayagam and M.~Rangamani,
  %``CFT dual of the AdS Dirichlet problem : Fluid/Gravity on cut-off surfaces,''
  JHEP {\bf 1112}, 090 (2011)
  [arXiv:1106.2577 [hep-th]].
  %%CITATION = ARXIV:1106.2577;%%
  %32 citations counted in INSPIRE as of 13 Feb 2014

%\cite{Camps:2010br}
\bibitem{Camps:2010br}
  J.~Camps, R.~Emparan and N.~Haddad,
  %``Black Brane Viscosity and the Gregory-Laflamme Instability,''
  JHEP {\bf 1005}, 042 (2010)
  [arXiv:1003.3636 [hep-th]].
  %%CITATION = ARXIV:1003.3636;%%
  %37 citations counted in INSPIRE as of 13 Jan 2014

%\cite{Emparan:2012be}
\bibitem{Emparan:2012be}
  R.~Emparan and M.~Martinez,
  %``Black Branes in a Box: Hydrodynamics, Stability, and Criticality,''
  JHEP {\bf 1207}, 120 (2012)
  [arXiv:1205.5646 [hep-th]];
  %%CITATION = ARXIV:1205.5646;%%
  %2 citations counted in INSPIRE as of 13 Jan 2014

%\cite{Emparan:2013ila}
\bibitem{Emparan:2013ila}
  R.~Emparan, V.~EHubeny and M.~Rangamani,
  %``Effective hydrodynamics of black D3-branes,''
  JHEP {\bf 1306}, 035 (2013)
  [arXiv:1303.3563 [hep-th]].
  %%CITATION = ARXIV:1303.3563;%%
  %7 citations counted in INSPIRE as of 13 Jan 2014



\bibitem{Bai:2012ci}
  X.~Bai, Y.~-P.~Hu, B.~-H.~Lee and Y.~-L.~Zhang,
  %``Holographic Charged Fluid with Anomalous Current at Finite Cutoff Surface in Einstein-Maxwell Gravity,''
  JHEP {\bf 1211}, 054 (2012)  [arXiv:1207.5309 [hep-th]].  %%CITATION = ARXIV:1207.5309;%%  %5 citations counted in INSPIRE as of 05 Jun 2013

%\cite{Hu:2013lua}
\bibitem{Hu:2013lua}
  Y.~-P.~Hu, Y.~Tian and X.~-N.~Wu,
  %``Bulk Viscosity of dual Fluid at Finite Cutoff Surface via Gravity/Fluid correspondence in Einstein-Maxwell Gravity,''
  Phys.\ Lett.\ B {\bf 732}, 298 (2014)
  [arXiv:1311.3891 [hep-th]];
  %%CITATION = ARXIV:1311.3891;%%
  %1 citations counted in INSPIRE as of 05 Aug 2014

%\cite{Hu:2014wka}
\bibitem{Hu:2014wka}
  Y.~-P.~Hu, Y.~Tian, X.~-N.~Wu, H.~-F.~Li and H.~-S.~Zhang,
  %``Breaking the apparent degeneracy between bulk viscosity and perturbation of the pressure in gravity/fluid correspondence,''
  arXiv:1408.2173 [hep-th].
  %%CITATION = ARXIV:1408.2173;%%


%\cite{Grozdanov:2011aa}
\bibitem{Grozdanov:2011aa}
  S.~Grozdanov,
  %``Wilsonian Renormalisation and the Exact Cut-Off Scale from Holographic Duality,''
  JHEP {\bf 1206}, 079 (2012)
  %doi:10.1007/JHEP06(2012)079
  [arXiv:1112.3356 [hep-th]].
  %%CITATION = doi:10.1007/JHEP06(2012)079;%%
  %9 citations counted in INSPIRE as of 16 May 2017




%%%%%%%%%%%%%%%%%%%%%%%%%%%%%%%%%%%%%%%%%%%%%%%%%%%%%%%%%Counterterm in EB gravity
%\cite{Balasubramanian:1999re,de Haro:2000xn,Emparan:1999pm}
%\cite{Balasubramanian:1999re}
\bibitem{Balasubramanian:1999re}
  V.~Balasubramanian and P.~Kraus,
  %``A stress tensor for anti-de Sitter gravity,''
  Commun.\ Math.\ Phys.\  {\bf 208}, 413 (1999)
  [arXiv:hep-th/9902121].
  %%CITATION = CMPHA,208,413;%%

%\cite{de Haro:2000xn}
\bibitem{de Haro:2000xn}
  S.~de Haro, S.~N.~Solodukhin and K.~Skenderis,
  %``Holographic reconstruction of space-time and renormalization in the AdS /
  %CFT correspondence,''
  Commun.\ Math.\ Phys.\  {\bf 217}, 595 (2001)
  [arXiv:hep-th/0002230].
  %%CITATION = CMPHA,217,595;%%


%\cite{Emparan:1999pm}
\bibitem{Emparan:1999pm}
  R.~Emparan, C.~V.~Johnson and R.~C.~Myers,
  %``Surface terms as counterterms in the AdS/CFT correspondence,''
  Phys.\ Rev.\  D {\bf 60}, 104001 (1999)
  [arXiv:hep-th/9903238].
  %%CITATION = PHRVA,D60,104001;%%


%%%%%%%%%%%%%%%%%%%%%%%%%%%%%%%%%%%%%%%%%%%%%%%%%%%%%%%%%%%%%%%%%%%%%%%%%%%%%%%Conformal transformation
%\cite{Myers:1999psa}
\bibitem{Myers:1999psa}
  R.~C.~Myers,
  %``Stress tensors and Casimir energies in the AdS / CFT correspondence,''
  Phys.\ Rev.\ D {\bf 60}, 046002 (1999)
  %doi:10.1103/PhysRevD.60.046002
  [hep-th/9903203].
  %%CITATION = doi:10.1103/PhysRevD.60.046002;%%
  %201 citations counted in INSPIRE as of 21 Jun 2017


%\cite{deHaro:2000vlm}
\bibitem{deHaro:2000vlm}
  S.~de Haro, S.~N.~Solodukhin and K.~Skenderis,
  %``Holographic reconstruction of space-time and renormalization in the AdS / CFT correspondence,''
  Commun.\ Math.\ Phys.\  {\bf 217}, 595 (2001)
  %doi:10.1007/s002200100381
  [hep-th/0002230];
  %%CITATION = doi:10.1007/s002200100381;%%
  %1001 citations counted in INSPIRE as of 21 Jun 2017
%\cite{Skenderis:2000in}
%\bibitem{Skenderis:2000in}
  K.~Skenderis,
  %``Asymptotically Anti-de Sitter space-times and their stress energy tensor,''
  Int.\ J.\ Mod.\ Phys.\ A {\bf 16}, 740 (2001)
  %doi:10.1142/S0217751X0100386X
  [hep-th/0010138].
  %%CITATION = doi:10.1142/S0217751X0100386X;%%
  %158 citations counted in INSPIRE as of 21 Jun 2017




\end{thebibliography}
\end{document}